\documentclass[twocolumn,twoside,slac_two]{revtex4}
\usepackage{graphicx}
\usepackage{fancyhdr}
\begin{document}

\title{Increasing the Number of TeV Blazars with Parsec-Scale Kinematics}

%

\author{Vivian C. Tiet, B. Glenn Piner}
\affiliation{Department of Physics and Astronomy, Whittier College,
13406 E. Philadelphia Street, Whittier, CA 90608, USA}
\author{Philip G. Edwards}
\affiliation{CSIRO Astronomy and Space Science,
PO Box 76, Epping NSW 1710, Australia}

\begin{abstract}
We report on our observations of the parsec-scale radio jet structures of five blazars 
that have been detected by ground-based TeV gamma-ray telescopes. 
These five blazars all belong to the class of High-frequency peaked BL Lac objects (HBLs), 
which are the most common blazar type detected at the TeV energy range. 
Because of their relative faintness in the radio, these HBLs are not well represented in other radio blazar surveys. 
Our observations consist of five epochs of Very Long Baseline Array (VLBA) imaging from 2006 to 2009, 
of each of the five blazars 1ES~1101$-$232, Markarian~180, 1ES~1218+304, PG~1553+113, and H~2356$-$309, 
at frequencies from 5 to 22~GHz. Fundamental jet properties, including the apparent jet speeds, 
that can be measured from these multi-epoch series of VLBA images are presented and compared with other gamma-ray blazars. 
Confirming prior work, we find that the TeV HBLs have significantly slower apparent jet speeds than
radio selected blazars. Together with other radio properties of the HBL class, this suggests modest
Lorentz factors in their parsec-scale radio jets. This in turn
suggests some form of Lorentz factor gradient in these jets,
since they are likely to have high Lorentz factors in their TeV-emitting regions.
The study presented here approximately doubles the number of TeV HBLs with multi-epoch parsec-scale kinematic measurements.
\end{abstract}

\maketitle

\thispagestyle{fancy}


\section{INTRODUCTION}
The new generation of ground-based TeV gamma-ray telescopes 
has now detected about 50 AGN as of this writing (tevcat.uchicago.edu). 
Most of these belong to the class of  
high-frequency peaked BL Lac objects (HBLs), with synchrotron and inverse Compton 
spectra peaking in the X-rays and high-energy gamma-rays, respectively. 
These blazars display remarkable variability in their gamma-ray emission on time scales 
of minutes (e.g.,~\cite{A07}), suggesting extremely high Lorentz factors 
of about 50~in the TeV emitting region of their relativistic jets (e.g.,~\cite{B08}).

The only way to image these relativistic jets directly on the parsec-scale is in the 
radio with VLBI. Most HBLs are relatively faint in the radio, so they are not 
included in other flux-limited VLBI monitoring projects. We are currently studying 
the parsec-scale structures of the TeV HBLs with multi-epoch VLBI imaging
using the National Radio Astronomy Observatory's Very Long Baseline Array (VLBA).
Our results for six of these TeV blazars have already been reported~\cite{P10}, 
here we present multi-epoch VLBA observations of an additional 
five TeV blazars discovered by the TeV telescopes during 2006.

Studying the parsec-scale structure of these radio-weaker high-frequency peaked
BL Lac objects with VLBI is fundamentally important to blazar physics, 
as they are likely to represent a physically distinct
AGN class from the flat-spectrum radio quasars and the low and intermediate-frequency
peaked BL Lac objects. It has been suggested that the former
group (the HBLs) represent non-evolving low-excitation radio galaxies (LERGs) with
less efficient or absent accretion disks and weak jets, while the latter
group (the FSRQs, LBLs, and IBLs) represent evolving high-excitation radio galaxies (HERGs)
with standard accretion disks and powerful jets~\cite{G12}.
It has also been suggested that unified schemes applied to the weak-jet sources 
are consistent with velocity gradients in the radiating plasma 
in the jets of these weak-jet sources~\cite{M11}, and such velocity gradients
may be directly observable in VLBI imaging in the form of transverse structures in the images.

\begin{figure*}
\centering
\includegraphics[scale=0.86]{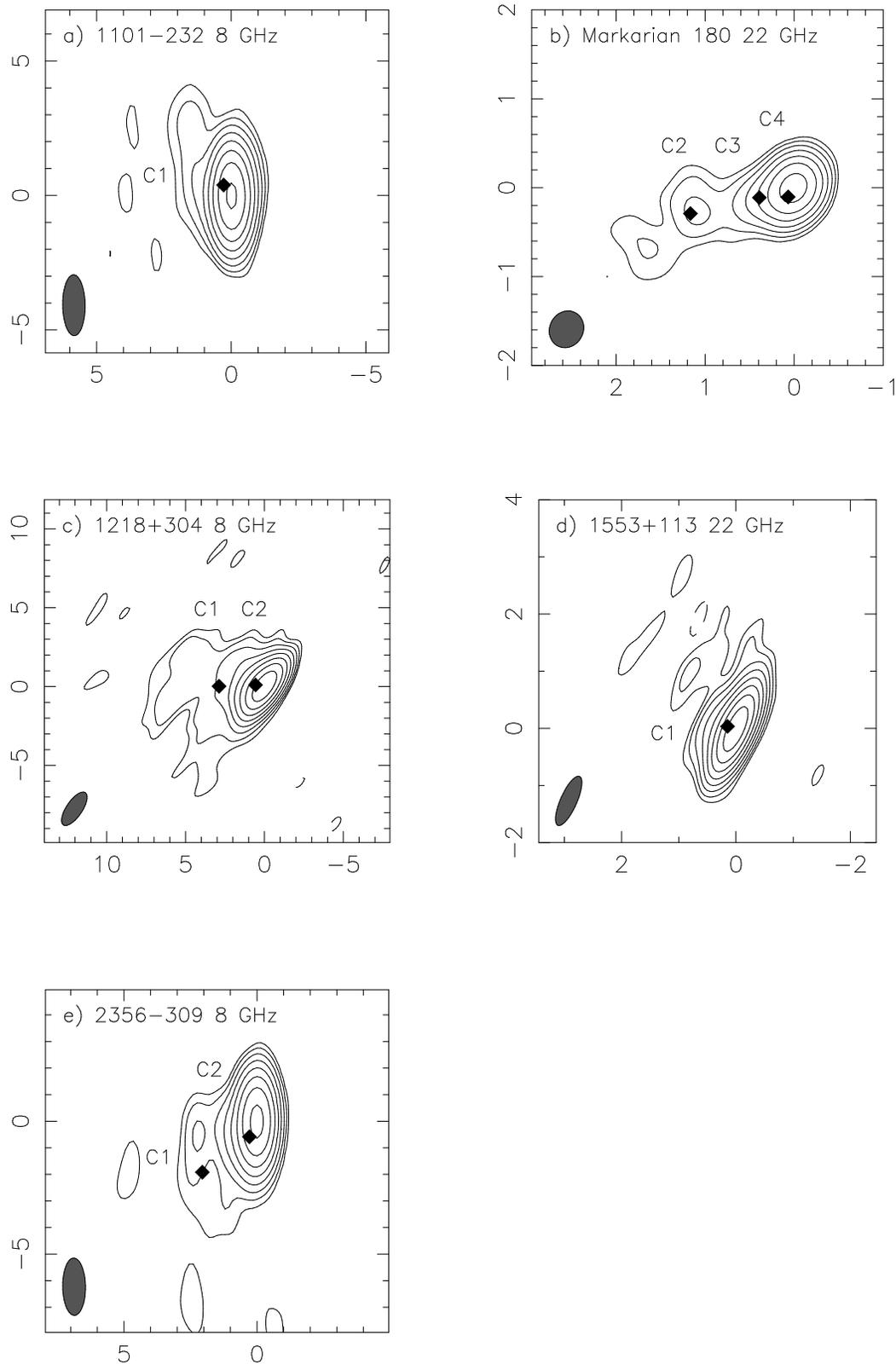}
\caption{VLBA images of 1ES~1101$-$232, Markarian~180, 1ES~1218+304, PG~1553+113, and H~2356$-$309
at 8 or 22~GHz from 2009 Jul 2. The axes are labeled in milliarcseconds (mas).
The lowest contour in each image is three times the rms noise level in that image, and each
successive contour is a factor of two higher.
Peak flux densities are 23, 60, 29, 91, and 22 mJy beam$^{-1}$ for Figures $a$ to $e$, respectively.
The rms noise levels are 0.05, 0.22, 0.04, 0.14, and 0.05 mJy beam$^{-1}$ for Figures $a$ to $e$, respectively.
Positions of Gaussian component centers are indicated by diamonds.}
\end{figure*}

\section{OBSERVATIONS}
We observed the five TeV blazars 1ES 1101-232, Markarian 180, 1ES 1218+304, PG 1553+113, 
and H 2356-309 at five epochs each from 2006 to 2009 with the VLBA under observation code BP146,
at frequencies from 5 to 22 GHz, for a total of 25 new VLBA images
produced for this paper. With the VLBA, observing at higher frequencies provides superior
angular resolution but worse sensitivity. For each source observed for this paper, the observation frequencies
were chosen to obtain the best possible angular resolution while still 
significantly detecting the parsec-scale jet.

Most observations recorded 
three hours of data on source, at a data rate of 256 Mbps. Polarization observations were 
made of Markarian 180 and PG 1553+113 (the two brightest sources) at the first epoch. 
The data were calibrated in AIPS, and imaged in Difmap. 
All sources were detected with direct fringe fitting at all epochs, with peak flux
densities ranging from 14 to 148 mJy~beam$^{-1}$. Phase referencing was not needed
for delay and rate corrections; although phase-referenced observations were made of 
the three sources 1ES~1101$-$232, 1ES~1218+304, and H~2356$-$309 at
the first epoch in order to determine more accurate milliarcsecond scale positions
of these sources than were previously available.

An image of each of these five blazars, from the 2009 Jul 2 epoch (the final epoch), is shown in Figure~1.
After imaging, we fit circular or elliptical Gaussian components to the calibrated visibilities 
using the {\em modelfit} task in Difmap.
Note that fitting to the visibilities allows sub-beam resolution to be obtained, so that
components are clearly separated from the core in the model fitting even when they appear 
blended with the core in the images. 
The complete set of 25 images and the corresponding models will be published elsewhere.
After model fitting, linear fits were made to the separations from the core of the Gaussian model component
centers versus time, in order to measure the apparent jet speeds.
For this fitting, error bars were assigned to the model component positions  
using the method described by~\cite{P07}.  

\begin{figure*}
\centering
\includegraphics[angle=90,scale=0.58]{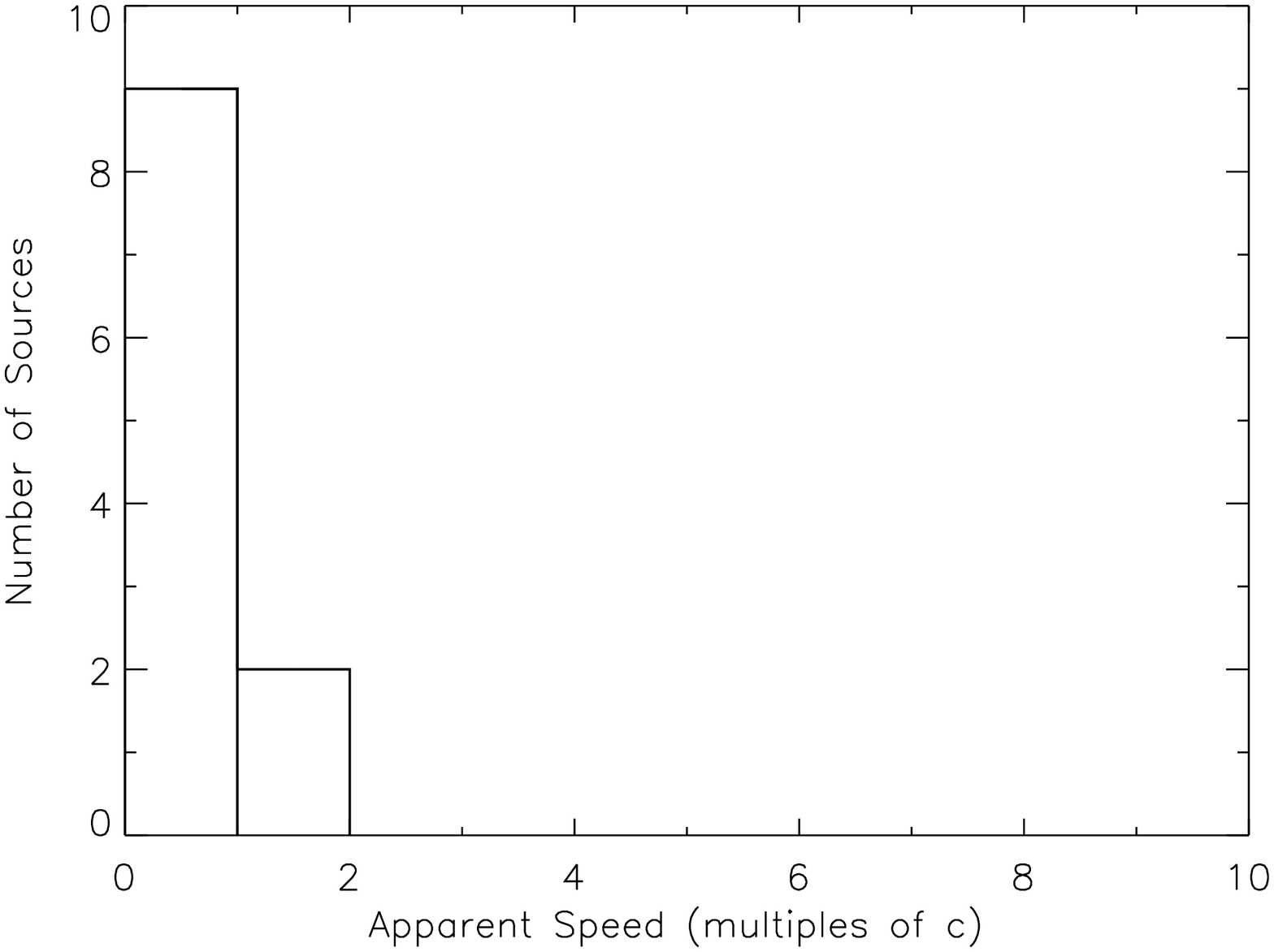}
\includegraphics[angle=90,scale=0.58]{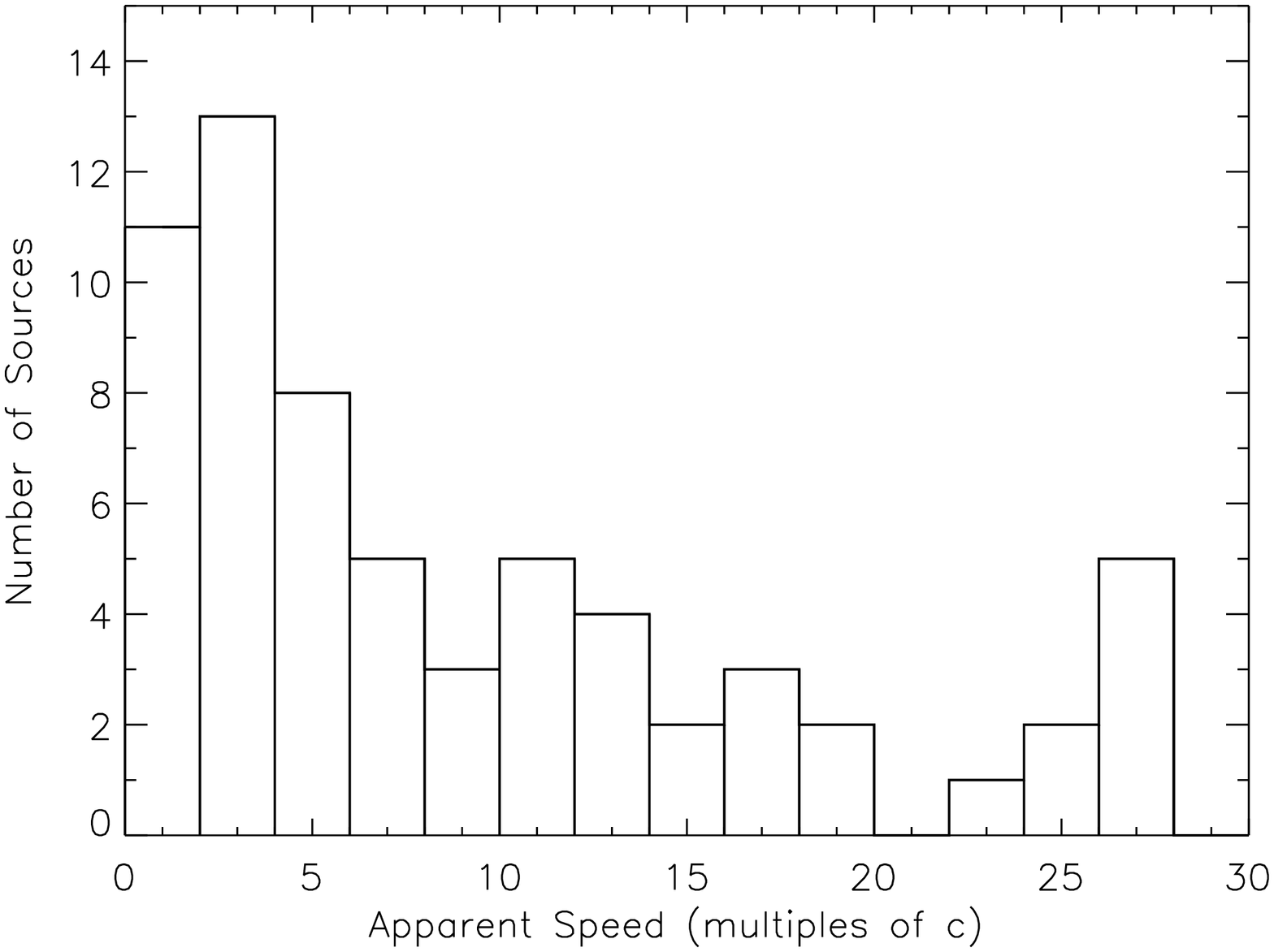}
\caption{Distribution of fastest apparent jet speeds in the 11 TeV HBLs studied by us to date (top), 
compared with the fastest apparent jet speeds measured in the radio-selected blazar sample 
from the Radio Reference Frame Image Database~\cite{P12} (bottom). 
All apparent speeds are measured from first order linear fits to 
the separations from the core of the Gaussian model component
centers versus time.
For the TeV HBLs, six of the speeds are from Table~5~in~\cite{P10},
while the remaining five come from this paper.
Formally negative speeds, most of which are statistically consistent with no motion,
are included in the leftmost bin.} 
\end{figure*}

\section{THE INDIVIDUAL SOURCES}
\subsection{1ES~1101$-$232}
Observations of this blazar at a redshift of $z$=0.186 were made at 
frequencies of 5 and 8~GHz.
Its structure consists of a 25~mJy core, and a short jet extension to the
northeast that is represented as a single 5~mJy component in our modeling.

\subsection{Markarian~180}
Markarian~180 (1133+704, $z$=0.045) is among the two brightest of these five blazars in total
radio flux density, and we observed it at a frequency of 22~GHz. 
At this frequency, a jet extends several mas to the east from a 40~mJy core.
This jet is modeled by four Gaussian components at most epochs.
VLBI polarization observations were done at the first epoch, and they showed that the VLBI core
is 4\% polarized, with the electric vector position angle aligned with the jet position angle.

\subsection{1ES~1218+304}
Observations of this $z$=0.182 blazar were made at frequencies of 5 and 8~GHz.
The images show a jet with a large apparent opening angle extending about 10~mas to the east from a 25~mJy core.
The inner 3~mas of this jet are modeled as two stationary Gaussian components in our observations.

\subsection{PG~1553+113}
The redshift of this source is somewhat uncertain.
The redshift has been constrained to be in the range from 0.43 to 0.58~\cite{D10},
and we have used the mean of this range, or $z$=0.5, for results requiring a redshift measurement. 
This source is relatively bright, and our VLBA observations were made at a frequency of 22~GHz. 
This source is quite compact at 22~GHz,
with a $\sim100$~mJy core and a short jet to the northeast that can be modeled as a single
$\sim25$~mJy component about 0.2~mas from the core.
A single epoch of polarization observations at the first epoch showed the core to be 3\% polarized, with the
electric vector position angle aligned with the jet position angle.

\subsection{H~2356$-$309}
Observations of the $z$=0.165 source H~2356$-$309 were made at 5 and 8~GHz.
H~2356$-$309 is both the faintest and the southernmost of the five blazars
observed in this series, making the imaging more difficult. The parsec-scale structure of this source consists of
a $\sim$20~mJy core and a jet extending several mas to the southeast
that is modeled by two components.

\begin{table}[t]
\begin{center}
\caption{Apparent Component Speeds}
\begin{tabular}{l c r r} \colrule \colrule \\
& & \multicolumn{1}{c}{Proper Motion} & \multicolumn{1}{c}{Apparent Speed} \\
\multicolumn{1}{c}{Source} & \multicolumn{1}{c}{Comp.} & \multicolumn{1}{c}{(mas yr$^{-1}$)}
& \multicolumn{1}{c}{(multiples of $c$)} \\ \colrule
1ES~1101$-$232 & C1 & $-0.173\pm0.129$ & $-2.07\pm1.54$ \\
Markarian~180  & C1 & $0.272\pm0.231$  & $0.81\pm0.69$  \\
               & C2 & $0.084\pm0.049$  & $0.25\pm0.15$  \\
               & C3 & $0.020\pm0.053$  & $0.06\pm0.16$  \\
               & C4 & $0.003\pm0.018$  & $0.01\pm0.05$  \\
1ES~1218+304   & C1 & $0.086\pm0.089$  & $1.01\pm1.05$  \\
               & C2 & $-0.067\pm0.050$ & $-0.78\pm0.58$ \\
PG~1553+113    & C1 & $0.032\pm0.036$  & $0.95\pm1.09$  \\
H~2356$-$309   & C1 & $-0.404\pm0.354$ & $-4.32\pm3.78$ \\
               & C2 & $-0.095\pm0.106$ & $-1.02\pm1.14$ \\ \colrule
\end{tabular}
\end{center}
\end{table}

\section{RESULTS}
For these conference proceedings, we present results on the apparent jet speed measurements in these five blazars.
Other results from this series of VLBA observations, including transverse structure measurements,
polarization images, brightness temperatures, and flux variability measurements
will be presented in a more complete paper elsewhere.

All together, we measured apparent speeds for 10 different jet components in these five blazars,
approximately doubling the number of TeV HBLs for which multi-epoch
parsec-scale kinematics information is available.
These new apparent speed measurements are tabulated in Table~1.
Apparent speed values were calculated assuming cosmological parameters of
$H_{0}=71$ km s$^{-1}$ Mpc$^{-1}$, $\Omega_{m}=0.27$, and $\Omega_{\Lambda}=0.73$.
Note that there is a large range in the proper motion error associated with the different components.
In general, the relatively bright compact components detected at all epochs in the high-frequency
observations have more precise proper motion measurements than the fainter more diffuse components detected
in the lower-frequency observations.

The apparent speeds presented in Table~1 are all low, 
with values ranging to only slightly over 1$c$, and the measurements 
actually show no detected motions at all that are above $2\sigma$ significance. 
This is in marked contrast to radio and GeV-selected blazar samples, 
which have typical apparent jet speeds of several $c$ (e.g.,~\cite{P07},~\cite{L09}).
The distributions of the fastest measured apparent jets speeds in each jet for the 
eleven TeV HBLs that we have observed to date and
for a radio-selected blazar sample are compared in Figure~2.
The distributions differ at $>$99.99\% confidence according to a Kolmogorov-Smirnov (KS) test.
This difference between the HBLs and the more powerful blazars
was previously noted for the first six TeV blazars studied 
with the VLBA~\cite{P10}, and now these additional five TeV HBLs
confirm and statistically strengthen that earlier result.

\section{DISCUSSION}
When considered together with several other lines of evidence discussed in detail
in~\cite{P10} and~\cite{P08}, this 
lack of superluminal apparent speeds in the TeV HBLs suggests that the Lorentz factor
is relatively low in the parsec-scale radio jets of these sources ($\Gamma\sim$2-3),
which contrasts with the high Lorentz factors that are typically required to explain their
TeV gamma-ray variability and spectral energy distribution.

These modest radio-jet Lorentz factors imply that the relativistic jets of the TeV HBLs
may be structured, with a Lorentz factor gradient either along the jet
(so that the jet decelerates by the parsec scale, e.g.,~\cite{G03})
or transverse to the jet (a fast spine surrounded by a slower layer or sheath, e.g.,
~\cite{G05}), or both. The existence of limb brightening
in the VLBI images of the nearest TeV blazars (Markarian 421 and 501,~\cite{P10},~\cite{P09},
~\cite{G04},~\cite{G06})
supplies additional direct evidence that transverse structures exist in these jets.
A number of other models in addition to simple Lorentz factor gradients have also been proposed
to explain the Lorentz factor differences between the radio and TeV-emitting regions in these sources,
see~\cite{P10} and references therein for a more complete discussion of models in the literature.

Of the additional newly discovered TeV HBLs that have been detected by the TeV telescopes within the past couple of years,
a number have quite low flux densities in the radio, below that of the blazars
that were imaged for this paper. However, the sensitivity upgrades recently 
completed at the VLBA should allow the parsec-scale structures of all of the
extragalactic TeV detections to be imaged in the near future, and definitively compared with
the more powerful sources detected by {\em Fermi} that are already well-studied in the radio.

\bigskip 
\begin{acknowledgments}
The National Radio Astronomy Observatory is a facility of the National Science Foundation
operated under cooperative agreement by Associated Universities, Inc.
This work was supported by the National Science Foundation under grant 0707523.
\end{acknowledgments}

\bigskip 

\end{document}